# Distinguishing a SM-like MSSM Higgs boson from SM Higgs boson at muon collider

Jai Kumar Singhal

*Department of physics*
*Government College, Sawai Madhopur 322 001 India*
*Email: jksinghal@hotmail.com; singhalph@sancharnet.in*

Sardar Singh and Ashok K Nagawat
*Department of Physics, University of Rajasthan, Jaipur 302 004 India*

## Abstract

We explore the possibility of distinguishing the SM-like MSSM Higgs boson from the SM Higgs boson via Higgs boson pair production at future muon collider. We study the behavior of the production cross section ($\sigma$) in SM and MSSM with Higgs boson mass ($m_{Higgs}$) for various choices of MSSM parameters $\tan\beta$ and $m_A$. We observe that at fixed $\sqrt{s}$, in the SM, the total cross section increases with the increase in Higgs boson mass whereas this trend is reversed for the MSSM case. The changes that occur for the MSSM case in comparison to the SM predictions are quantified in terms of the relative percentage deviation in cross section ($\Delta\sigma\%$). We find that for $m_A = 100$ GeV (at $\sqrt{s} = 500$ GeV, $\tan\beta = 5 - 50$), $\Delta\sigma\%$ is negative and its magnitude increases with increase in $m_{Higgs}$ whereas, for $m_A = 300$ GeV the $\Delta\sigma\%$ is positive (at $m_{Higgs} = 115$ GeV) and reduces to zero and then becoming negative with increasing $m_{Higgs}$. We also find that for given $\sqrt{s}$, $m_A$ and $m_{Higgs}$ the variation in $\Delta\sigma\%$ is more prominent in lower $\tan\beta$ region in comparison to higher $\tan\beta$ region. The observed large deviations in cross section for different choices of $m_{Higgs}$ suggest that the measurements of the cross section could possibly distinguish the SM-like MSSM Higgs boson from the SM Higgs boson. Furthermore, the dependence of $\Delta\sigma\%$ on the MSSM parameters $\tan\beta$ and $m_A$ for the given $m_{Higgs}$ indicates that information on $\tan\beta$ and $m_A$ could be obtained from the cross section measurements.

**Keywords:** Higgs boson; SM; MSSM; SM like MSSM Higgs boson.

**PACS No.** 14.80.-j; 14.8o.Bn; 14.80.Cp

# I. INTRODUCTION

The existence of the scalar Higgs boson (*H*) of the Standard Model (SM) is still waiting for experimental confirmation [1, 2]. The direct searches for the SM Higgs boson at the LEP II have achieved a 95% CL bound of $m_H > 114.4$ GeV [2]. The global fits to all precision electroweak data give $m_H = 113^{+56}_{-40}$ and $m_H < 241$ GeV (95% CL) [3]. It has been argued that if such a Higgs boson exists, it fits more naturally into the Minimal Supersymmetric Standard Model (MSSM) then into SM itself [4, 5]. Moreover, a Higgs boson with mass ~115 GeV in the context of the supersymmetry would mess nicely with the evidence of anomalous magnetic moment of muon [3, 5].

In many extensions of the SM the Higgs sector is enlarged, containing several neutral Higgs bosons as well as charged ones [6]. In particular, the MSSM predicts three neutral (*h*, $H^0$ and *A*) and two charged ($H^\pm$) Higgs bosons [6]. An important prediction of the MSSM is an upper bound to the mass of the lightest CP – even Higgs boson (*h*). At tree level, one obtains $m_h \leq m_Z |\cos 2\beta|$. Such a light Higgs particle (*h*) is essentially ruled out by the searches at LEPII [1]. However, once radiative corrections are incorporated the theoretical upper bound on $m_h$ is raised substantially and is restricted to $m_h < 135$ GeV [7-9]. The MSSM possesses a limit, called decoupling limit that is experimentally almost indistinguishable from the SM [10]. This occurs when the pseudo scalar mass is large (i.e., $m_A \gg m_Z$), then the CP-even ($H^0$), CP-odd ($A^0$) and charged ($H^\pm$) Higgs bosons are mass degenerate and the mass of the lightest CP-even Higgs boson (*h*) approaches to its upper bound value for the given tan *β*. In this limit the lightest MSSM Higgs boson (*h*) and the SM Higgs boson (*H*) have very similar properties, i.e., in this limit the SM-like MSSM Higgs boson mimics the signature of the SM Higgs boson, and therefore even if a neutral scalar boson is discovered in the near future, the task of discriminating between SM-like MSSM Higgs boson and SM Higgs boson will be quite hard [11].

It is hoped that at least one Higgs boson within the mass range allowed by the MSSM will be discovered at Tevatron and/or LHC [12]. It has been argued that precision measurements of the properties of the Higgs sector at a linear collider may allow one to discriminate between SM and SM-like MSSM Higgs bosons and further extract or constrain the model parameters [13]. Therefore, it is of the interest to explore the possibility of distinguishing between these two particles.

In recent years an increasing amount of work has been dedicated to the physics possibilities of $\mu^+\mu^-$ colliders [14-19]. It has been suggested that a muon collider might prove essential to the understanding of the Higgs sector of a SUSY model by accurately measuring the properties of a light SM-like Higgs boson and distinguishing it from a supersymmetric Higgs boson [5].

The process $e^+e^- \rightarrow$ neutral Higgs boson pair is not of much interest due to smallness of *Hee* couplings. However, the amplitude for process $\mu^+\mu^- \rightarrow$ neutral Higgs boson pair, at the tree-level is enhanced by a factor ~ $m_\mu/m_e$. At tree level, in (i) SM and (ii) MSSM, the process $\mu^+\mu^- \rightarrow$ neutral Higgs boson pair occurs via *s*-channel and *t*-channel diagrams shown in Fig. 1 and 2 respectively. The contribution of one extra *s*-channel diagram in MSSM and corresponding interference in $|amplitude|^2$ may open a possibility of distinguishing the SM-like MSSM Higgs boson from the SM Higgs boson at a muon collider. In view of this, we study the neutral Higgs boson pair production at muon collider.

This paper is organized as follows: In Sec. II we perform the calculations for amplitudes and cross sections for the processes $\mu^+\mu^- \to HH$ and $\mu^+\mu^- \to hh$ where $H$ is SM Higgs boson and $h$ is SM like MSSM Higgs boson. In Sec. III we examine the behavior of cross section ($\sigma$) for the production of (i) SM Higgs boson pair and (ii) SM-like MSSM Higgs boson pair, and (iii) the relative percentage deviation $\Delta\sigma\% = \left(\dfrac{\sigma_{MSSM} - \sigma_{SM}}{\sigma_{SM}}\right) \times 100\%$. We summarize our conclusions in Sec. IV.

## II. DETAILS OF THE CALCULATIONS

For sake of completeness, we first define the kinematics of the reaction $\mu^+\mu^- \to HH$ ($hh$). Let $k$ ($\bar{k}$) and $\sigma(\bar{\sigma})$ are the four momenta and helicities of incoming $\mu^-$ ($\mu^+$) respectively and $q(\bar{q})$ is the four momenta of the outgoing Higgs particles. The momenta in the centre of mass frame of $\mu^+\mu^-$ system are given by (here $\sqrt{s}$ = total c. m. energy and we have assumed $\sqrt{s} \gg m_\mu$):

$$k = \frac{\sqrt{s}}{2}[1, 0, 0, 1], \qquad q = \frac{\sqrt{s}}{2}[1, \beta\sin\theta, 0, \beta\cos\theta],$$

$$\bar{k} = \frac{\sqrt{s}}{2}[1, 0, 0, -1], \qquad \bar{q} = \frac{\sqrt{s}}{2}[1, -\beta\sin\theta, 0, -\beta\cos\theta],$$

where $\theta$ is the scattering angle of $H(h)$ with respect to the incident $\mu^-$ direction, $\beta = \sqrt{1 - \frac{4m_{H(h)}^2}{s}}$, $\gamma = \frac{\sqrt{s}}{2m_{H(h)}}$ and $m_{H(h)}$ is the SM Higgs boson (SM like MSSM Higgs boson) mass.

## A. THE PROCESS $\mu^+\mu^- \rightarrow HH$

In the SM the process $\mu^+(k,\sigma) + \mu^-(\bar{k},\bar{\sigma}) \rightarrow H(q) + H(\bar{q})$ at tree level proceeds via $H$–exchange in $s$-channel and $\mu$– exchange in $t$– and crossed $t$–channels (Fig.1). The s-channel $\gamma$- and $Z$- exchange are forbidden by CP-invariance [20].

The relevant SM couplings (in unitary gauge) are [21]

$H\mu^+\mu^-$ :  $-\frac{igm_\mu}{2m_W}$ $\qquad\qquad$ $HHH$ :  $-\frac{3igm_H^2}{2m_W}$.

We evaluate the amplitude of the process $\mu^+\mu^- \rightarrow HH$ following the technique described in detail by Renard [22]. We find that the amplitude for the process is

$$M_{SM}(\sigma,\bar{\sigma}) = 3\sqrt{2s}\, G_F m_\mu \left[ \frac{m_H^2}{s - m_H^2} \delta_{\sigma,\sigma} - \frac{8\sigma m_\mu \beta^2 \sin\theta \cos\theta}{3\sqrt{s}\left(\frac{1}{\gamma^4} + 4\beta^2 \sin^2\theta\right)} \delta_{\sigma,-\sigma} \right]. \quad (1)$$

The differential cross section for the process in case of unpolarized initial muon beams is then given by

$$\frac{d\sigma_{SM}}{d\cos\theta} = \frac{1}{2} \frac{\beta}{128\pi s} \sum_{\sigma,\bar{\sigma}} |M_{SM}(\sigma,\bar{\sigma})|^2. \quad (2)$$

Here a factor of ½ is taken because of two identical particles in the final state (see appendix B of the Ref. [21]). Using Eq. (1) in Eq. (2), we obtain

$$\frac{d\sigma_{SM}}{d\cos\theta} = \frac{9G_F^2 m_\mu^2 \beta}{64\pi}\left[\left(\frac{m_H^2}{s-m_H^2}\right)^2 + \left(\frac{8m_\mu \beta^2 \sin\theta\cos\theta}{3\sqrt{s}\left(1/\gamma^4 + 4\beta^2\sin^2\theta\right)}\right)^2\right] \quad (3)$$

The total cross section is obtained by integrating the above and is found to be

$$\sigma_{SM} = \frac{9G_F^2 m_\mu^2 \beta}{32\pi}\left[\left(\frac{m_H^2}{s-m_H^2}\right)^2 + \frac{8m_\mu^2\beta^2}{9s}\left\{\begin{array}{l}4\gamma^4 + \dfrac{9}{4\beta^2} + \dfrac{1}{2\beta\xi} \\ \times\left(1+\dfrac{3}{8\gamma^4\beta^2}\right)\log\left(\dfrac{\xi+2\beta}{\xi-2\beta}\right)\end{array}\right\}\right], \quad (4)$$

with
$$\xi = \sqrt{\frac{1}{\gamma^4} + 4\beta^2} \ . \quad (5)$$

## B. THE PROCESS $\mu^+\mu^- \to hh$

In the MSSM the Higgs sector is enlarged to contain two scalar doublet fields (one coupling to up- and other coupling to down-type fermions and whose non zero expectation values induce spontaneous electroweak symmetry breaking), leading to five Higgs particles: two CP-even ($h$ and $H^0$), a CP odd ($A$) and two charged ($H^\pm$) Higgs bosons. The supersymmetric structure of the theory imposes constraints on the Higgs sector of the model, as a result, all Higgs sector parameters at tree-level are determined by just two free parameters; conventionally chosen as the ratio of vacuum expectation values of each doublet ($tan\beta = v_2/v_1$) and mass of CP-odd Higgs boson ($m_A$). At tree level, one obtains $m_h \leq m_Z|\cos 2\beta|$. This bound is modified by radiative corrections and restricted to $m_h <$ 135 GeV [7-9].

Now we evaluate the amplitude for the process $\mu^+(k,\sigma) + \mu^-(\bar{k},\bar{\sigma}) \to h(q) + h(\bar{q})$. The contributions to the process arise due to $s$- channel $h$- and $H^0$-exchange and $\mu$-exchange in $t$- and crossed $t$-channel diagrams (see Fig. 2). The Bose symmetry forbids the *Zhh*-vertex [23]. Below we summarize the couplings needed for our study [23]:

$$h\mu^+\mu^-: \quad \frac{-igm_\mu}{2m_W}\lambda_{h\mu^+\mu^-}, \qquad H^0\mu^+\mu^-: \quad \frac{-igm_\mu}{2m_W}\lambda_{H^0\mu^+\mu^-},$$

$$hhh: \quad \frac{-3ig}{2m_W}m_Z^2 \cos 2\alpha \sin(\alpha+\beta),$$

$$H^0hh: \quad \frac{-ig}{2m_W}m_Z^2 \left[2\sin 2\alpha \sin(\alpha+\beta) - \cos 2\alpha \cos(\alpha+\beta)\right],$$

with $\quad \lambda_{h\mu^+\mu^-} = -\dfrac{\sin\alpha}{\cos\beta} \quad$ and $\quad \lambda_{H^0\mu^+\mu^-} = \dfrac{\cos\alpha}{\cos\beta}.$

The amplitude for the process is found to be

$$M_{MSSM}(\sigma,\overline{\sigma}) = -3\sqrt{2s}\, G_F m_\mu \left[\left(\frac{a}{s-m_h^2} - \frac{b}{s-m_{H^0}^2}\right)\delta_{\sigma,\sigma} + \frac{8\sigma m_\mu \beta^2}{3\sqrt{s}} \\ \times \frac{\sin\theta\cos\theta}{(1/\gamma^4 + 4\beta^2 \sin^2\theta)}\left(\frac{\sin\alpha}{\cos\beta}\right)^2 \delta_{\sigma,-\sigma}\right] \quad (6)$$

where $\quad a = m_Z^2 \dfrac{\sin\alpha \cos 2\alpha \, \sin(\alpha+\beta)}{\cos\beta} \quad (7)$

$$b = m_Z^2 \frac{\cos\alpha[2\sin 2\alpha \sin(\alpha+\beta) - \cos 2\alpha \cos(\alpha+\beta)]}{3\cos\beta}. \quad (8)$$

Here α is the mixing angle that rotates the weak CP-even Higgs eigenstates into the mass eigenstates *h* and *H*[0] and is given by [24]

$$\tan 2\alpha = \tan 2\beta \frac{(m_A^2 + m_Z^2)}{\left(m_A^2 - m_Z^2 + \dfrac{\varepsilon}{\cos 2\beta}\right)} \quad (9)$$

with $\quad \varepsilon = \dfrac{3G_F}{\sqrt{2}\,\pi^2}\dfrac{m_{top}^4}{\sin^2\beta}\log\left(1 + \dfrac{m_S^2}{m_{top}^2}\right), \quad (10)$

where $m_{top}$ and $m_S$ are top and s-top quark masses.

The differential cross section is

$$\frac{d\sigma_{MSSM}}{d\cos\theta} = \frac{9G_F^2 m_\mu^2 \beta}{64\pi}\left[\left(\frac{a}{s-m_h^2}-\frac{b}{s-m_{H^0}^2}\right)^2 + \left(\frac{\sin\alpha}{\cos\beta}\right)^4 \times \left(\frac{8m_\mu\beta^2 \sin\theta\cos\theta}{3\sqrt{s}(1/\gamma^4 + 4\beta^2 \sin^2\theta)}\right)^2\right], \quad (11)$$

and the total cross section is found to be

$$\sigma_{MSSM} = \frac{9G_F^2 m_\mu^2 \beta}{32\pi}\left[\left(\frac{a}{s-m_H^2}-\frac{b}{s-m_{H^0}^2}\right)^2 + \left(\frac{\sin\alpha}{\cos\beta}\right)^4 \frac{8m_\mu^2\beta^2}{9s} \times \left\{4\gamma^4 + \frac{9}{4\beta^2} + \frac{1}{2\beta\xi}\left(1+\frac{3}{8\gamma^4\beta^2}\right)\log\left(\frac{\xi+2\beta}{\xi-2\beta}\right)\right\}\right]. \quad (12)$$

## III. BEHAVIOUR OF CROSS SECTIONS AND $\Delta\sigma\%$

For numerical evaluation, we note that the limit from direct searches at LEP II in the MSSM context excludes $m_h < 91.0$ GeV and $m_A < 91.9$ GeV at 95% CL [1] and $m_h$ is theoretically restricted to be < 135 GeV with the inclusion of radiative corrections. For SM Higgs boson the current experimental bound is $m_H$ >114.4 GeV, therefore we take the Higgs boson mass ($m_{Higgs} = m_{H(h)}$) ranging from 115 to 135 GeV. For MSSM parameters we choose $m_{top}$= 175 GeV and $m_{stop}$ = 1 TeV. We use: $G_F$ =1.16637×10$^{-5}$ GeV$^{-2}$, $m_\mu$=105.658 MeV, $m_Z$ = 91.1876 GeV.

In Fig. 3 we plot the total cross section as a function of Higgs boson mass for the production of Higgs boson pair in the SM (solid line) and in the MSSM (dashed and dotted lines), for tan$\beta$ = 5 to 50 and $m_A$ = 100 to 300 GeV. We note that in the case of SM higgs boson pair production cross section increases with the increase in the Higgs boson mass. On the other hand for the SM like MSSM Higgs boson pair production the cross section decreases with the increase of the Higgs boson mass. Furthermore, for a given Higgs boson mass the cross section depends on the choice of the $\tan\beta$ and $m_A$.

In order to quantify the changes that occurs for the SM-like MSSM Higgs boson pair production in comparison to SM Higgs boson pair production we define the relative percentage deviation in the cross section by the relation

$$\Delta\sigma\% = \left(\frac{\sigma_{MSSM} - \sigma_{SM}}{\sigma_{SM}}\right) \times 100\% \ . \tag{13}$$

It is a measure of deviation from the SM predictions for the production of neutral Higgs boson pairs in the $\mu^+\mu^-$ annihilation.

The behavior of the $\Delta\sigma\%$ with Higgs boson mass for $\tan\beta = 5$ (solid line), 25 (dashed line) and 50 (dotted line) and $m_A = 100$ and 300 GeV at fixed $\sqrt{s} = 500$ GeV is shown in Fig. 4. The value of $\Delta\sigma\%$ depends on $m_A$ and $\tan\beta$ for the given Higgs boson mass. It is found that for $m_A = 100$ GeV, the $\Delta\sigma\%$ is negative and the magnitude increases with increase in Higgs boson mass. For $m_A = 300$ GeV, we note that the $\Delta\sigma\%$ is positive at $m_{Higgs} = 115$ GeV and the magnitude decreases with increasing Higgs boson mass. The $\Delta\sigma\%$ depends significantly on the values of $m_{Higgs}$, $\tan\beta$ and $m_A$.

The variation of $\Delta\sigma\%$ as a function of $\tan\beta$ is displayed in Fig. 5 for $m_A = 100$, 200 and 300 GeV and $m_{Higgs} = 115$ (solid line) and 135 GeV (dashed line) at fixed $\sqrt{s} = 500$ GeV. It demonstrates the strong dependence on choice of $m_{Higgs}$ and $m_A$ for a given $\tan\beta$. Further, we note that for given set of $\sqrt{s}$, $m_A$ and $m_{Higgs}$ the variation in $\Delta\sigma$ % is more prominent in lower tan$\beta$ region in comparison to higher tan$\beta$ region.

## IV. CONCLUSIONS AND DISCUSSION

We have considered the pair production of (i) SM Higgs bosons and (ii) SM-like MSSM Higgs bosons in $\mu^+\mu^-$ collisions. We have examined pair production cross section ($\sigma$) and relative percentage deviation ($\Delta\sigma$ %) for various values of $m_{Higgs}$, and MSSM parameters tan$\beta$ and $m_A$. Our conclusions are:

(i) At fixed $\sqrt{s}$ = 500 GeV, the total cross section ($\sigma$) increases with the increase in Higgs boson mass in the case of SM Higgs boson pair production. However, for SM like MSSM Higgs boson pair production the cross section ($\sigma$) decreases with the increase of Higgs boson mass.

(ii) To examine the changes that occur for the SM-like MSSM Higgs boson pair production in comparison to SM Higgs boson pair production we defined the relative percentage deviation in cross section ($\Delta\sigma$ %). We observe the following:

    (a) At $\sqrt{s}$ = 500 GeV and $m_A$ = 100 GeV (tan$\beta$ = 5 to 50), $\Delta\sigma$ % is negative and its magnitude increases with the increase in Higgs boson mass.

    (b) At $\sqrt{s}$ = 500 GeV and $m_A$ = 300 GeV (tan$\beta$ = 5 to 50), $\Delta\sigma$ % is positive (for $m_{Higgs}$ = 115 GeV) and decreases with increasing

Higgs boson mass. It approaches zero (e.g., at $\tan\beta = 5$ and $m_{Higgs}$ = 130 GeV) and then becomes negative.

(c) For fixed $\sqrt{s}$, $m_A$ and Higgs boson mass the variations in $\Delta\sigma\%$ are more prominent in low $\tan\beta$ region (e.g., $\Delta\sigma\% = -80.05\%$ and $-25.36\%$ for $\tan\beta = 2.5$ and 10 respectively at $\sqrt{s} = 500$ GeV, $m_A = 100$ GeV and $m_{Higgs} = 115$ GeV) in comparison to large $\tan\beta$ region (e.g., $\Delta\sigma\% = -05.21\%$ and $-03.06\%$ for $\tan\beta = 30$ and 50 respectively at $\sqrt{s} = 500$ GeV, $m_A = 100$ GeV and $m_{Higgs} = 115$ GeV).

(iii) The observed large deviation in total cross section for different choices of Higgs boson masses indicates that the measurements of the cross section could possibly distinguish the SM-like MSSM Higgs boson from the SM Higgs boson.

(iv) For the any choice of the Higgs boson mass the $\Delta\sigma\%$ strongly depends on the choice of the MSSM parameters $\tan\beta$ and $m_A$, as such the information on $\tan\beta$ and $m_A$ could be obtained from the cross section measurements.

In arriving the above conclusions we have considered the SM Higgs boson pair production and SM- like MSSM Higgs boson production at tree level. However, these processes also occur at loop level. The process $\mu^+\mu^- \to HH$ at one loop level is mediated only by W and Z loops, while in MSSM, additional contributions to the corresponding process $\mu^+\mu^- \to hh$ will originate from chargino, neutralino, s- muon, s-neutrino loops as well as loops built up by the associated $A$ and $H^{\pm}$ bosons. In this regard we note that the influence of supersymmetric particles on the precision electroweak measurements is generally negligible [25], since radiative corrections mediated by SUSY particles are suppressed by a factor of $\frac{m_Z^2}{m_S^2}$, where $m_S$ is the scale characterizing the scale of the SUSY particles. For example, loop induced pair production of SM Higgs boson and SM-like MSSM Higgs boson in $e^+e^-$ collisions has been considered in literature [20] and it was found that for $hh$ production, contribution of SUSY loops are in general rather small: in fact at high energies the SUSY boxes practically do not contribute; but at low energies, and specially below the decoupling limit, the SUSY contribution can be of the order of ~10%, and maximum contribution of SUSY loops (for some parameter space) was found to be $\sim -15\%$ [20]. In the decoupling limit, the SUSY contributions are at the most of the order of few percent and the cross sections are therefore of the same order as in the SM and deviation from the SM prediction is small at one loop level [20]. As such we expect that the inclusion of radiative corrections will not substantially change our conclusions.

# REFERENCES

[1]   S Eidelman *et al.* (Review of Particle Properties), Phys. Lett. **B592**, 1 (2004).

[2]   ALEPH, DELPHI, L3 and OPAL Collaborations and the LEP Working Group for Higgs Boson Searches, Phys. Lett. **B565**, 61 (2003).

[3]   J Erler and P Langacker, in Ref. [1] section 10; W J Marciano, *Precision Electroweak Measurements and the Higgs Mass*, hep-ph/ 0411179.

[4]   T Abe *et al.*, *Linear Collider Physics Resource Book for Snowmass 2001 Part 2: Higgs and Supersymmetry Studies*, hep-ex/ 0106056 (2001).

[5]   V Barger, M S Berger, J F Gunion and T Han, in Proc. of the *APS/ DPF/DPB Summer study on the future of particle physics* (Snowmass 2001), *eds*, R Davidson and C Quigg, [hep-ph/0110340].

[6]   For reviews, see, M Carena and H E Haber, Prog. Part. Nucl. Phys. **50**, 63 (2003); J F Gunion, H E Haber, G Kane and S Dawson, *The Higgs Hunters Guide*, (Addison-Wesley, 1990).

[7]   M S Berger, Phys. Rev. **D 41**, 225 (1990); H E Haber and R Hempfling, Phys. Rev. Lett. **66**, 1815 (1991); Y Okada, M Yamaguchi and T Yanagida, Prg. Theor. Phys. **85**, 1 (1991); J Ellis, G Ridolfi and F Zwirner, Phys. Lett. **B 257**, 83 (1991).

[8]   One of the effects of radiative corrections to Higgs sector of the MSSM is the modification of the upper bound of lightest CP-even Higgs boson mass, as noted in Ref. [7]

[9]   The radiative corrections have been computed by a number of techniques and a variety of approximations at one and two loops. For an exhaustive list of references and discussion see, M Carena, J S Conway, H E Haber and J Hobes *et al.*, *Report of the Tevatron Higgs Working Group*, hep-ph/ 0010338.

[10]  H E Haber and Y Nir, Phys. Lett. **B306**, 327(1993); H E Haber, *in Physics from the Planck scale to the Electroweak scale*, Proc. of the US-Polish

Workshop, Warsaw, Sept. 21-24, 1994, *eds*. P Nath, T Taylor and S Pokorski (World Scientific, 1995) p. 49.

[11] A Dobado, M J Herrero and S Penaranda, Eur. Phys. J. **C17**, 487 (2000).

[12] For review and references see, P Igo-Kemenes, *Searches for Higgs Bosons*, in Ref. [1]; J F Gunion, H E Haber and R V Kooten in *Linear Collider Physics in the new Millennium*, *eds*. K Fujii, D Miller and A Soni (World Scientific): hep-ph/0301023; S Dawson and M Oreglia, Ann. Rev. Nucl. Part. Sci. **54**, 269 (2004).

[13] M Carena, H E Haber, H E Logan and S Mrenna, Phys. Rev. **D65**, 055005 (2002); Erratum *ibid* **D65**, 099902(20002); S Dawson, S Heinemeyer and S Mrenna, Phys. Rev. **D66**, 055002 (2002); J Guasch, W Hollik and S Penaranda, Phys.Lett. **B515**, 367 (2001); K Desch, T Klimkovich, T Kuhl and A Raspereza, *Study of Higgs Boson Pair Production at Linear Collider*, hep-ph/0406229

[14] M M Alsharo'a *et al*., Phys. Rev. ST. Accel. Beams **6**, 081001 (2003); D. B. Cline, J. Phys. **G29**, 1661 (2003); I Bigi *et al*, Phys. Rep. **371**, 151 (2002); C Blochinger *et al*., *Physics Opportunities at mu+mu- Higgs Factories*, Report of the Higgs factory working group of the ECFA-CERN study on Neutrino Factory and Muon Storage Rings at CERN, hep-ph/0202199; R Raja *et al*., *The program in muon and neutrino physics: Super beams, Cold muon beams, neutrino factory and the muon collider*, Submitted to Snowmass 2001, hep-ex/0108041; C. M. Ankenbrandt *et al*., Muon Collider Collaboration, Phys. Rev. ST. Accel. Beams **2**, 08100 (1999).

[15] V Barger, M S Berger, J F Gunion and T Han, Phys. Rev. Lett. **75**, 1462 (1995); *ibid*. **78**, 3991 (1997); Phys. Rep. **286**, 1 (1997).

[16] M S Berger, *Precision W-Boson and Higgs Boson Mass Determination at Muon Colliders*, hep-ph/9712474 (1997); *Threshold Cross-Section Measurements*, hep-ph / 9802213 (1998); *Muon Collider Physics at Very High Energies*, hep-ph/0001018 (2000); *SUSY Thresholds at a Muon Collider*, hep-ph/0003128 (2000).

[17] B Kamal, W Marciano and Z Parsa, in the Proc. of the *Workshop on Physics at the First Muon Collider and at the Front End of the Muon Collider* (Fermilab, Nov. 1997), edited by S Geer and R Raja, AIP Conf. Proc. **435**, (1998) p. 657.

[18] J K Singhal, and Sardar Singh, Phys. Rev. **D64**, 013007 (2001); J K Singhal, Sardar Singh, A K Nagawat and N K Sharma, Phys. Rev. **D63**, 017302 (2001)

[19] P E Asakawa, S Y Choi and J S Lee, Phys. Rev. **D63**, 015012 (2001); V Barger, T Han and C G Zhou, Phys. Lett. **B480**, 140 (2000); A G Akeroyd, A Arhrib and C Dove, **D61**, 071724R (2000); B Grzadkowski, J F Gunion and J Pliszka, Nucl. Phys. **B583**, 49 (2000); G J Gunaris and F M Renard, Phys. Rev. **D59**, 113015 (1999).

[20] A Djouadi, V Driesen and C Junger, Phys. Rev. **D54**, 759 (1996).

[21] C Quigg, *Gauge theories of the strong, weak and electromagnetic interactions* (The Benjamin/Cumming publishing company, Inc.1983), p. 129.

[22] F. M. Renard, **Basics of electron positron collisions**, Edition Frontieres, Gif sur Yvette, France (1981).

[23] J F Gunion and H E Haber, Nucl. Phys. **B272**, 1 (1986); Erratum hep-ph / 9301205.

[24] see for example, M Spira, Fortschr. Phys. **46**, 203 (1998).

[25] H E Haber, *Where are radiative corrections important in the minimal supersymmetric model?*, hep-ph/9305248.

# Figure Captions

Fig. 1. Tree level Feynman diagrams for the process $\mu^+\mu^- \to HH$. The relevant couplings are mentioned in Sec. II.

Fig. 2. Tree level Feynman diagrams for the process $\mu^+\mu^- \to hh$. The relevant couplings are mentioned in Sec. II.

Fig.3. Variation of total cross section ($\sigma$) with Higgs boson mass ($m_{Higgs}$) for different values of $\tan\beta$ and $m_A$ at fixed $\sqrt{s}$ = 500GeV. The solid line is for SM. The dashed and dotted lines are for MSSM.

Fig.4. Behavior of $\Delta\sigma\%$ with Higgs boson mass ($m_{Higgs}$) for different values of $\tan\beta$ and $m_A$ (in GeV) at fixed $\sqrt{s}$ = 500GeV. The solid lines are for $\tan\beta$ = 5, dashed lines are for $\tan\beta$ = 25 and dotted lines are for $\tan\beta$ = 50.

Fig.5. Behavior of $\Delta\sigma\%$ as a function of $\tan\beta$ for different values of $m_{Higgs}$ and $m_A$ at fixed $\sqrt{s}$ = 500GeV. The solid lines are for $m_{Higg}$ = 115 GeV, dashed lines are $m_{Higg}$ = 135 GeV.

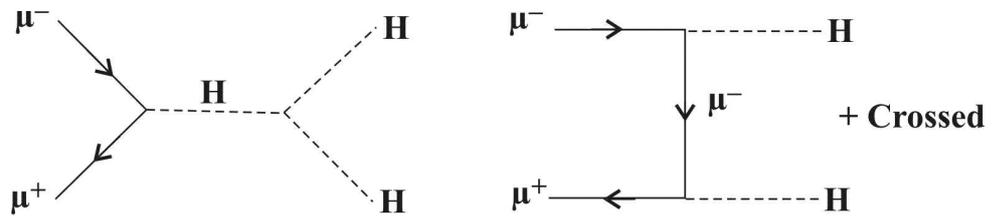

**Figure : 1**

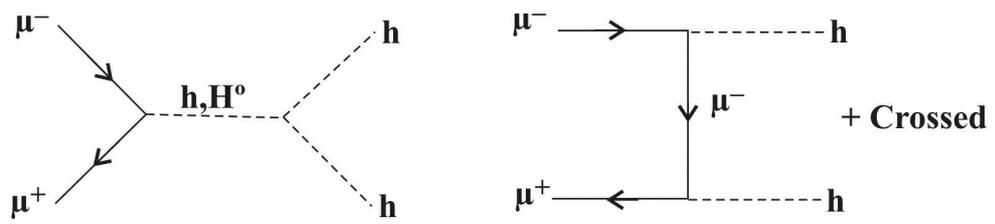

**Figure : 2**

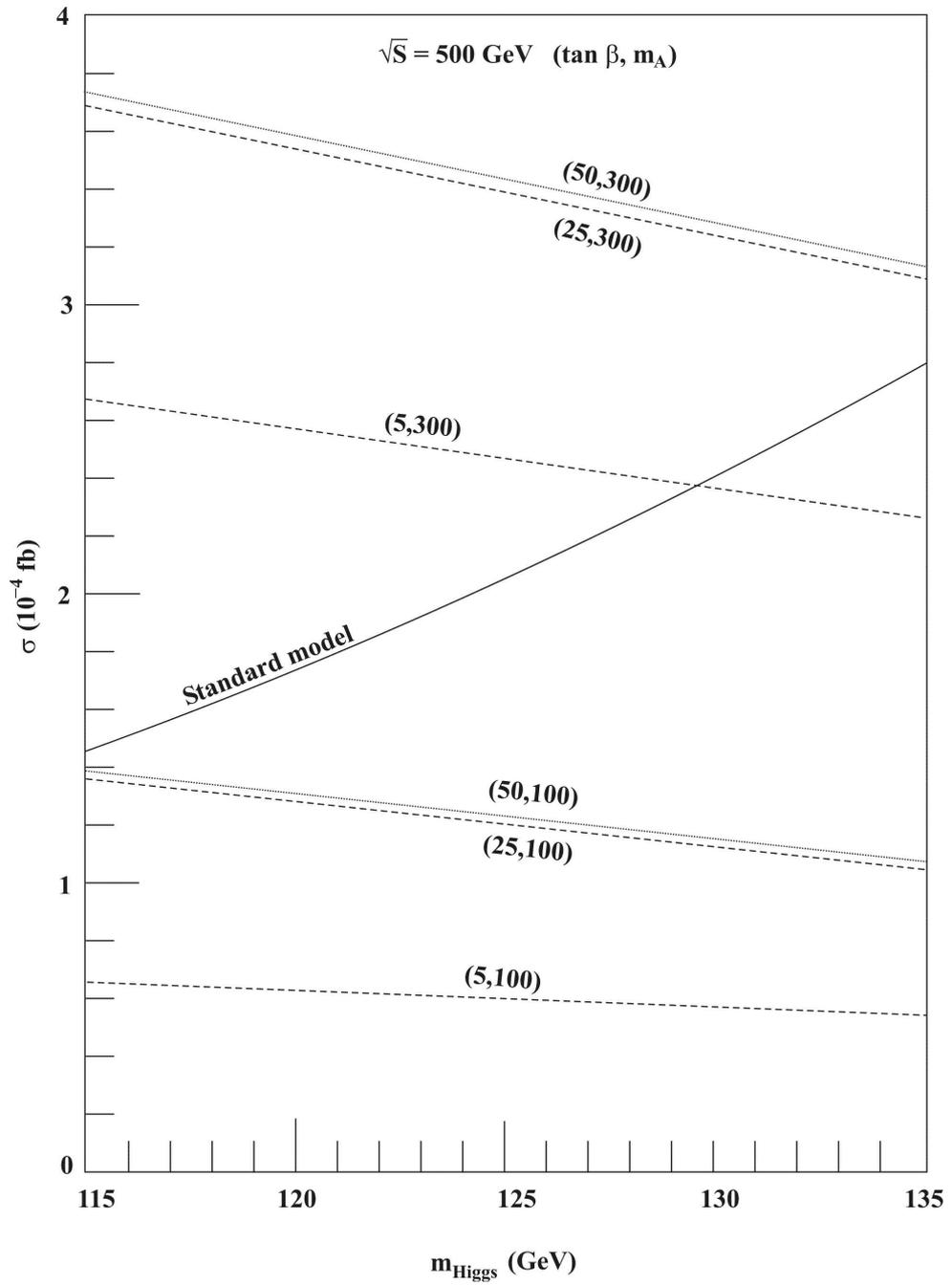

**Figure : 3**

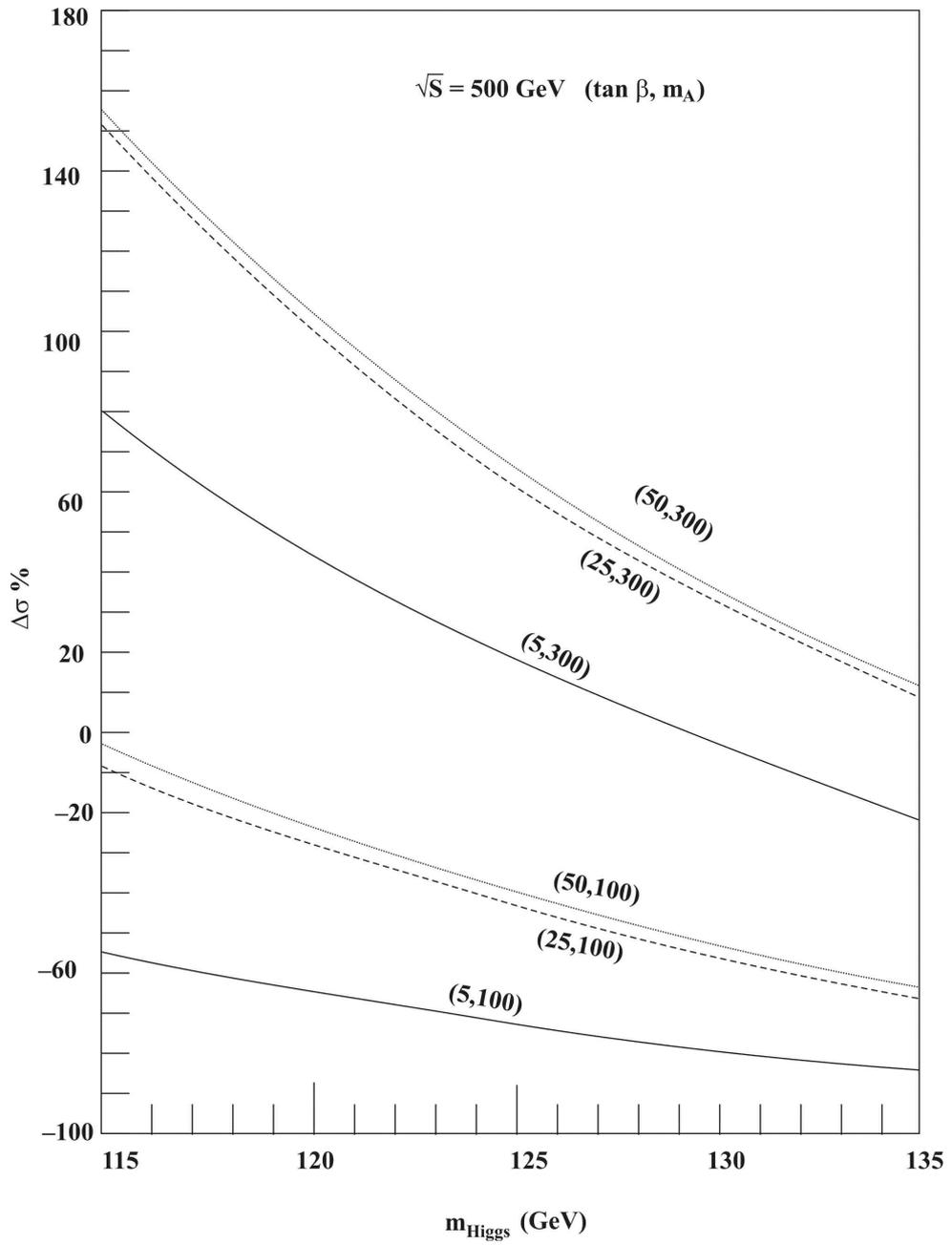

**Figure : 4**

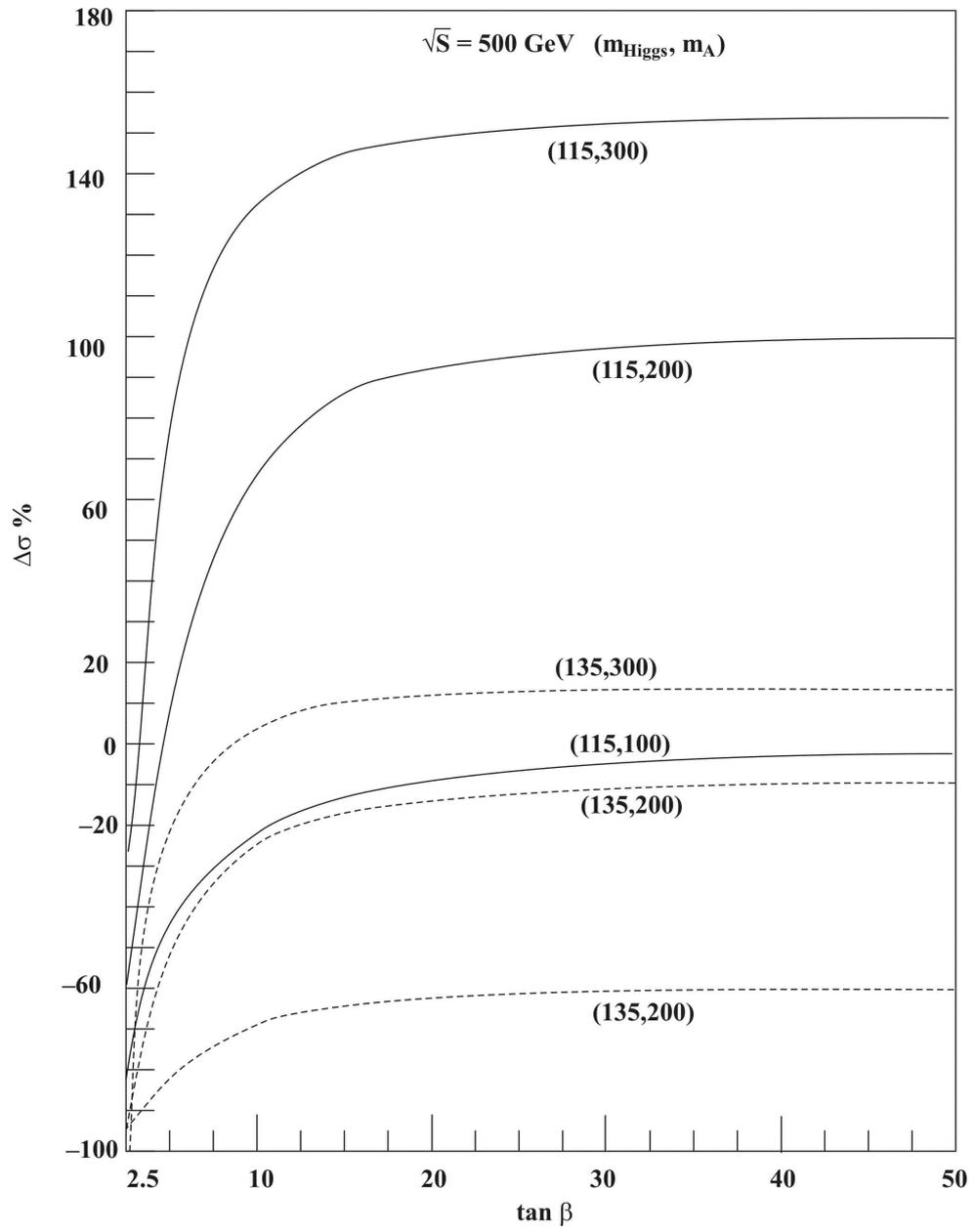

**Figure : 5**